\documentclass[twocolumn,aps,showpacs]{revtex4}
\newcommand{\be}{\begin{equation}}
\newcommand{\ee}{\end{equation}}
\newcommand{\ben}{\begin{eqnarray}}
\newcommand{\een}{\end{eqnarray}}

\newcommand{\bb}{\bibitem}
\begin{document}


\title{Topological defects and the trial orbit method}
\author{D. Bazeia, W. Freire, L. Losano, and R.F. Ribeiro}
\affiliation{Departamento de F\'\i sica, Universidade Federal da Para\'\i ba,
Caixa Postal 5008, 58051-970 Jo\~ao Pessoa, Para\'\i ba, Brazil}


\begin{abstract}
We deal with the presence of topological defects in models for
two real scalar fields. We comment on defects hosting topological defects,
and we search for explicit defect solutions using the trial orbit method.
As we know, under certain circunstances the second order equations of motion
can be solved by solutions of first order differential equations. In this
case we show that the trial orbit method can be used very efficiently to
obtain explicit solutions.
\end{abstract}
\maketitle

In the seventies there appeared a great deal of investigations concerning
the presence of topological defects and their role in high energy
physics. Particularly interesting issues appeared in the search for
defects in models involving real scalar fields, as for instance in the
investigations of Refs.~{\cite{rw75,m76,r79}} which considered models
described by two real scalar fields. The interest has
been renewed in the nineties, where there has appeared investigations
dealing with systems of two scalar field having distinct motivations,
as for instance in the case of defects springing in the form of junctions
of defects \cite{j1,j2,j3}, and also as defects having internal
structure \cite{i1,i2,i3}. Other lines of investigations concern the
presence of doman walls in supergravity \cite{sg1,sg2}, in scenarios
for localization of gravity on domain walls \cite{rs},
in supersymmetric gluodynamics, where nonperturbative
effects may give rise to gluino condensates \cite{g1,g2},
and also in string theory, since there are models in field theory which
correctly describe the low energy world volume dynamics of branes in string
theory \cite{s1,s2,s3}.

A central issue in the investigation of defects in systems
involving two real scalar fields concerns integrability of the equations
of motion. From the mathematical point of view the problem is
hard, because one starts with two coupled second order nonlinear
ordinary differential equations -- see Refs.~{\cite{rw75,m76,r79}}
for more details. However, in the investigations in Ref.~{\cite{95}}
one proposes a new route, in which the mathematical barrier is simplified
if one considers a specific class of systems. In this class of systems
the equations of motion can be reduced to first order differential equations,
which allows obtaining Bogomol'nyi-Prasad-Sommerfield (BPS)
states \cite{b,ps}, which are stable configurations that
minimize the energy of the topological solutions.

More recently, in Refs.~{\cite{01,es}} one gets two new results,
one showing that under certain conditions \cite{01} the second order
equations of motion can be reduced to a family of first order equations,
in this case ensuring that all the topological solutions are BPS states,
and the other showing that sometimes \cite{es} it is possible to find
an integrating factor for the first order equations which allows
obtaining all the BPS states of the system, thus unveiling the moduli
space of topological kinks.

As one knows, models described by two real scalar fields are generically
described by a Lagrange density, which contains the usual kinetic and
gradient contributions, and is otherwise specified by a potential, in
general a smooth function of the two scalar fields. If the
potential has several minima, one may find defects of at least
two different types: {\it topological defects}, which starts in
a given minimum of the potential and ends in another, distinct minimum;
{\it nontopological defects}, which starts and ends in the very same
minimum of the potential. These defects constitute orbits in the
plane $(\phi,\chi)$, an orbit being a function of the two fields
which can be represented by $F(\phi,\chi)=0$. In Ref.~{\cite{r79}}
the author has proposed a method to help searching for explicit solutions
to the equations of motion. The procedure has been named {\it trial orbit
method,} since it relies on trying an orbit $F(\phi,\chi)=0$, and checking
its validity. Although the method may help finding explicit solutions,
it is not that strong since it is directly related to solving the equations
of motion, which are second order differential equations. In the present work,
however, we revisit the trial orbit method, adapting the methodology to the
investigation of BPS states.

This possibility appears very naturally when one considers a specific class
of models described by two real scalar fields. We follow recent work
\cite{j3} and references therein to introduce models defined by
\be
\label{cm}
V(\phi,\chi)=\frac12\,W^2_{\phi}+\frac12\,W^2_{\chi}
\ee
Here $W=W(\phi,\chi)$ is a smooth function of the two fields, and
$W_{\phi}=\partial W/\partial\phi$ and so forth. In this case the
energy density for static field configurations $\phi(x)$ and $\chi(x)$
can be written in the form
\be
{\cal E}(x)=\frac{dW}{dx}+\frac12\left(\frac{d\phi}{dx}-W_{\phi}\right)^2+
\frac12\left(\frac{d\chi}{dx}-W_{\chi}\right)^2
\ee
and for the static field configurations that obey the first order
differential equations
\be
\label{f1}
\frac{d\phi}{dx}=W_{\phi},\;\;\;\;\;
\frac{d\chi}{dx}=W_{\chi}
\ee
the energy is minimized to the bound $E_{BPS}=|\Delta W|$, where
$\Delta W=W[\phi(\infty),\chi(\infty)]-W[\phi(-\infty),\chi(-\infty)]$
These solutions are BPS states, which solve the first order differential
equations and the equations of motion of the model.

To search for defect solutions we first investigate the minima
of the potential, which are critical points of $W$ obtained by
requiring that $W_{\phi}=0$ and
$W_{\chi}=0$. We suppose the model has the discrete set of $n$ minima,
$(\phi_{_1},\chi_{_1}), (\phi_{_2},\chi_{_2}),...,(\phi_{_n},\chi_{_n})$. 
In principle, each pair of minima forms a topological sector, so we use the
subscripts $ij$ to identify the topological sectors. Some of the topological
sectors may have solutions that obey the first order equations
(\ref{f1}). In this case they are named BPS sectors,
and we can identify the BPS sectors with the energy of the topological
sector, with $E_{ij}=|\Delta W_{ij}|$, for
$\Delta W_{ij}=W(\phi_i,\chi_i)-W(\phi_j,\chi_j)$.

We eliminate the spatial coordinate in the first order equations
$(\ref{f1})$ to write $W_{\chi}\;d\phi-W_{\phi}\;d\chi=0$.
This equation is exactly solvable for $W$ harmonic, for
$W_{\phi\phi}+W_{\chi\chi}=0$. In this case the solution is
given by the orbit $F(\phi,\chi)=0$, which is obtained via
$\partial F/\partial\phi=W_{\chi}$ and $\partial F/\partial\chi=-W_{\phi}$.
In Ref.~{\cite{01}} one has obtained that when $W$ is harmonic
all the topological solutions are of the BPS type, that is, are
solutions of the first order Bogomol'nyi equations. We use this
result together with the above result to introduce a new result,
which ensures that the equations of motion for topological solutions
are exactly solved when the potential is of the form
(\ref{cm}), with $W$ harmonic.

If $W$ is not harmonic, we can yet search for some integrating factor,
$I=I(\phi,\chi)$, such that the solution is now given by
$\widetilde F(\phi,\chi)=0$, with
$\partial\widetilde F/\partial\phi=I(\phi,\chi)\;W_{\chi}$
and
$\partial\widetilde F/\partial\chi=-I(\phi,\chi)\;W_{\phi}$.
The problem is that it is not always ease to find an integrating factor
for the first order equation. This difficulty has led us to revisit the
trial orbit method, to see if it can be of some help in the process
of finding solutions to the first order equations. As we will show,
the trial orbit method is of good use for searching for topological
defects in the BPS sectors that appear from the first order Bogomol'nyi
equations. Better than that, the trial orbit method is more efficient
than it appears in the original work \cite{r79}, since here it relies
on searching for solutions of first order equations, circumventing the
weakness that appears in Ref.~{\cite{r79}}, in which one deals with the
equations of motion, which are second order differential equations.

The trial orbit method may be seen as a procedure based on the
three steps:

{\bf Step 1.} We sellect the BPS sector. We do that by supposing the
pair of minima $(\phi_{_i},\chi_{_i})$ and
$(\phi_{_j},\chi_{_j})$ is such that
$W(\phi_{_i},\chi_{_i})\neq W(\phi_{_j},\chi_{_j})$, implying that one is
dealing with a BPS sector. This means that the first order equations
has topological solutions connecting the points $(\phi_{_i},\chi_{_i})$ and
$(\phi_{_j},\chi_{_j})$;

{\bf Step 2.} We choose an orbit. We do that by writing the equation of
the orbit, say $F_{ij}(\phi,\chi)=0$, and checking compatibility between
the chosen orbit and the minima that specify the BPS sector, that is,
validating the statements $F_{ij}(\phi_{_i},\chi_{_i})=0$ and
$F_{ij}(\phi_{_j},\chi_{_j})=0$;

{\bf Step 3.} We test compatibility between the orbit and the first order
equations. We do that by differentiating the trial orbit. We get
$(\partial F_{ij}/\partial\phi)(d\phi/dx)+
(\partial F_{ij}/\partial\chi)(d\chi/dx)=0.$
We use the first order equations (\ref{f1}) to obtain
\be
\label{ns}
\frac{\partial F_{ij}}{\partial\phi}\; W_{\phi}+
\frac{\partial F_{ij}}{\partial\chi}\; W_{\chi}=0
\ee
This new statement is similar to the trial orbit itself. Thus,
we have to check compatible between the chosen orbit
and the new statement (\ref{ns}). We say that $F_{ij}(\phi,\chi)=0$
is a good orbit when every pair $(\phi,\chi)$ that solves the
orbit also solves the new statement $(\ref{ns})$. We then use the
good orbit to solve the first order equations.

We examplify the procedure with some specific investigations. For simplicity
we deal with natural units, and with dimensionless fields and coordinates.
The first example is given by
\be
\label{m1}
W(\phi,\chi)=\phi-\frac13\phi^3-r\phi\chi^2
\ee
where the parameter $r$ is real. This is the model first investigated
in Ref.~{\cite{95}}. We use Eq.~(\ref{m1}) to write the first order equations
\ben
\label{m1e1}
\frac{d\phi}{dx}&=&1-\phi^2-r\chi^2
\\
\label{m1e2}
\frac{d\chi}{dx}&=&-2r\phi\chi
\een
We first consider $r$ real and negative. The
potential has minima at $v_{1,2}=(\pm1,0)$, which define a topological sector
of the BPS type. An explicit solution describes a straight line orbit
connecting the two minima; it corresponds to $\phi(x)=\tanh(x)$
and $\chi=0$. For $r=-1$ we see that $W$ is harmonic, and the model can
be solved exactly. There is no orbit connecting the minima $(\pm1,0)$,
but the single straight line segment with $\chi=0$. This result is valid
{\it not only} for the first order equations, {\it but also} for the
equations of motion, that is, the system has no topological solutions
connecting the minima $(\pm1,0)$ for $r=-1$, unless the kinks with
$\phi(x)=\pm\tanh(x)$ and $\chi=0$.
 
We now consider $r$ real and positive. In this case one can find
an integrating factor and exactly solve the model \cite{es}. However,
since we want to show how the trial orbit method works we proceed
using the methodology explained above. The minima of the potential are
now at $v_{1,2}(\pm1,0)$ and $v_{3,4}=(0,\pm\sqrt{1/r})$. The model has
one BPS sector with energy $E_{12}=4/3$, and four BPS sectors with energy
$E_{13}=E_{14}=E_{23}=E_{24}=2/3$.
To find non trivial solutions we use the above procedure. We follow the
first step to sellect the BPS sector defined by the minima $(\pm1,0)$, with
energy $E_{12}=4/3$. The second step leads us
to choose the orbit. We try $1+a\phi^2+b\,\chi^2=0$.
The ending points $(\pm1,0)$ demand that $1+a=0$, so we can write
$1-\phi^2+b\chi^2=0$. We follow
the third step and we differentiate this orbit to get
$-\phi(d\phi/dx)+b\chi(d\chi/dx)=0$.
We use the first order equations to get
$1-\phi^2+r(2b-1)\chi^2=0$.
This new statement is fully compatible with the orbit for
$b=r(2b-1)$, that is, for $b=r/(2r-1)$. This
finally gives the good orbit
\be
\label{o1}
\phi^2+\frac{r}{1-2r}\chi^2=1
\ee
which connects the minima $(\pm1,0)$ for $0<r<1/2$.

We now use the orbit $(\ref{o1})$ to find explicit solutions. It
allows rewriting Eq.~(\ref{f1}) in the form
$d\phi/dx=2r\,(1-\phi^2)$, which is solved by $\phi(x)=\tanh(2rx)$.
The other field is then given by
\be
\chi(x)=\pm\,\frac{\sqrt{(1-2r)/r}}{\cosh(2rx)}
\ee
We notice that the limit $r\to1/2$ leads to $\chi=0$ and
$\phi(x)=\tanh(x)$, 
which also solve the first order equations (\ref{f1}).
This pair corresponds to a straight line segment connecting
the minima $(\pm1,0)$. It is different from the other pair, which connects
the same minima by an elliptic segment. The presence of the elliptic orbit
manifests the possibility of domain walls having internal structure.
This possibility emerges after examining the masses of the two fields,
which are: $m^2_{\phi}=4,\,m^2_{\chi}=4r^2$ at the minima $(\pm1,0)$,
and $m^2_{\phi}=4r,\,m^2_{\chi}=4r$ at $(0,\pm\sqrt{1/r})$.
Also, the energies of the topological defects are given by: for $\chi=0$
we have $E_{12}=4/3$ and for $\phi=0$ we have $E_{34}=4/3\sqrt{r}$.
We see that for $r\in(0,1)$ energy considerations favor the BPS defect
to be the host defect, and for $r>1$ the non BPS defect becomes the
host defect. In both cases the elementary $\phi$ mesons
prefer to live inside the host defect, while the $\chi$ mesons prefer
to live outside.

In the other BPS sectors we can consider orbits like $r\,\chi^2=1\pm\phi$,
which requires $r=1/4$. For $r\,\chi^2=1-\phi$ we have
$2r\chi(d\chi/dx)=-d\phi/dx$. In this case we use Eqs.~(\ref{f1}) to obtain
$d\phi/dx=\phi-\phi^2$, which is solved by $\phi(x)=(1/2)[1+\tanh(x/2)]$.
The other field is given by $\chi(x)=\pm\sqrt{2[1-\tanh(x/2)]}$.
For $r\chi^2=1+\phi$ the investigation is similar. We get
$\phi(x)=-(1/2)[1-\tanh(x/2)]$ and
$\chi(x)=\pm\sqrt{2[1+\tanh(x/2)]}$.

In order to further illustrate our procedure we consider other
models, and we explore the most interesting BPS sectors they have.
The first model is defined by
\be
\label{m2}
W(\phi,\chi)=\phi-\frac13\, (1+s)\,\phi^3+
\frac15\, s\, \phi^5-\,r\,\phi\,\chi^2
\ee
It can be seen as an extension of the former model, which includes
the fifth order power on $\phi$. We suppose the parameters $r$ and $s$
are real. The first order equations are
\ben
\label{m2fo1}
\frac{d\phi}{dx}&=&1-(1+s)\,\phi^2+s\,\phi^4-r\,\chi^2
\\
\label{m2fo2}
\frac{d\chi}{dx}&=&-2\,r\,\phi\,\chi
\een
The potential $V(\phi,\chi)$ is given by
\ben
V(\phi,\chi)&=&\frac12(\phi^2-1)^2(s\phi^2-1)^2+\frac12 r^2\chi^4\nonumber\\
& &-r(1-\phi^2)(1-s\phi^2)\chi^2+
+2r^2\phi^2\chi^2
\een
It can be projected in the $\chi=0$ direction to give
\be
V(\phi,0)=\frac12 (\phi^2-1)^2(s\phi^2-1)^2
\ee
It is of the eight order power in $\phi$ and admits symmetry breaking
for $s$ positive and negative. It can also be
projected in the $\chi=0$ direction; in this case it gives
\be
V(0,\chi)=\frac12 (r\chi^2-1)^2
\ee
which is of the fourth order power in $\chi$ and admits symmetry breaking
for $r$ positive. The model contains the pair of minima $v_{1,2}=(\pm1,0)$
and also $v_{3,4}=(\pm\sqrt{1/s},0)$ if $s$ is positive, and also
$v_{5,6}=(0,\pm\sqrt{1/r})$ if $r$ is positive.

We use the potential to calculate the mass matrix, and to write the
masses of the two fields at the diverse minima of the potential. They are:
$m^2_{\phi}=4(1-s)^2,\,m^2_{\chi}=4r^2$ at the minima $(\pm1,0)$,
$m^2_{\phi}=4(1-s)^2/s,\,m^2_{\chi}=4r^2/s $ at $(\pm\sqrt{1/s},0)$,
and $m^2_{\phi}=4r,\,m^2_{\chi}=4r$ at $(0,\pm\sqrt{1/r})$.
This model admits the presence of domain walls having internal structure.
We examine this possibility in the case $r>0$ and $s<0$, where we have
the same four minima of the former model. Here, for $\chi=0$, in the
BPS sector defined by the minima $v_{1,2}=(\pm1,0)$ the energy is given by
$(4/3)(1+|s|/5)$; for $\phi=0$, in the non BPS sector
defined by $v_{3,4}=(0,\pm\sqrt{1/r})$ the energy is $4/3\sqrt{r}$.
Thus, the ratio $(5+|s|)/5\sqrt{r}$ shows that one favors
the presence of the BPS defect with $\chi=0$ as the host defect
for $\sqrt{r}>1+|s|/5$, and for $\sqrt{r}<1+|s|/5$ the host
defect is the non BPS defect with $\phi=0$. Thus, for $r\in(0,1)$
and for $s<0$ the host defect is necessarily the non BPS defect
that connects the minima $(0,\pm\sqrt{1/r})$. And in this case
the $\phi$ meson prefers to live outside the host defect, while
the $\chi$ meson prefers to live inside. This situation is similar
to the one that appears in the former model. However, for $r>1$
and for $s<0$, if one considers $\sqrt{r}>1+|s|/5$ the host defect
becomes the BPS defect that connects the minima $(\pm1,0)$. In this case,
the $\chi$ mesons prefer to live inside the host defect. But for the $\phi$
mesons we now have two possibilities: they prefer to live outside
the host defect if $\sqrt{r}>1+|s|$, and for $1+|s|/5<\sqrt{r}<1+|s|$
they also prefer to live inside the host defect. The present model
is more general than the former one, giving rise to the case where
the host defect entraps both the $\phi$ and $\chi$ mesons.

We notice that $W$ in Eq.~(\ref{m2}) is not harmonic, so the model is not
exactly solvable. Moreover, since we have been unable to find an integrating
factor for Eqs.~(\ref{m2fo1}) and (\ref{m2fo2}), we could not be sure that
the BPS states of the model can be solved exactly. For this reason,
let us now use the trial orbit method to explore the presence of BPS
states in this model. Because this new model contains an extra term, of the
fourth order power in $\phi$, we try the orbit
$a+b\phi^2+c\chi^2+d\phi^4=0$. We follow the former steps to get to
the good orbit $r\chi^2=s(1-\phi^2)^2$, for solutions connecting the
minima $(\pm1,0)$, with energy $(4/3)|1-s/5|$. We use this orbit
to rewrite Eq.~(\ref{m2fo1}) as
$d\phi/dx=(1-s)(1-\phi^2)$, which is solved by $\phi(x)=\tanh[(1-s)x]$.
We use this orbit to get for the other field the solutions
\be
\chi(x)=\pm\frac{\sqrt{s/r}}{{\cosh}^2[(1-s)x]}
\ee
which requires that $s/r>0$.
This pair of solutions exists for $r$ and $s$
positive or negative, so it appears in the sector connecting the minima
$(\pm1,0)$ despite the presence of the other minima. It goes to the solution
which describes a straight line orbit in the former model in the limit $s\to0$.

Let us now suppose that $s$ is positive, so that there are minima
at $v_{5,6}=(\pm\sqrt{1/s},0)$. In this case we can find BPS solutions
connecting those minima, with energy $(4/3\sqrt{s})|1-1/5s|$.
We follow the same procedure to find the orbit $r\chi^2=(1-s\phi^2)^2/s$.
This orbit allows obtaining the solutions
\ben
\phi(x)&=&\sqrt{1/s\,}\tanh\bigl[\sqrt{1/s\,}\;(s-1)x\bigr]
\\
\chi(x)&=&\pm\frac{\sqrt{1/rs\,}}{\cosh^2
\bigl[\sqrt{1/s}\;(s-1)x\bigr]}
\een
which requires $r>0$, so that there are minima also at $(0,\pm\sqrt{1/r})$.
This last solution appears if both $r$ and $s$ are positive, so that the
model must contains all the six minima $(\pm1,0)$, $(\pm\sqrt{1/s})$, and
$(0,\pm\sqrt{1/r})$.

We further illustrate the trial orbit method investigatimg another
model, defined by
\be
\label{m3}
W(\phi,\chi)=\phi-\frac13\,\phi^3-\frac12\,r\,\frac{\chi^2}{\phi}-
\frac{s}{\phi}+\frac13\frac{s}{\phi^3}
\ee
where $r$ and $s$ are real and positive, $s\in(0,1)$. The presence of
interactions leading to negative power in the fields is not unusual. They
appear in several different contexts, for instance in vortices in planar
Abelian systems involving generalized permeabilities \cite{v1,v2}, in effective
Yang-Mills theories coupled to scalar field, with appropriate color dielectric
function to mimic quark and antiquark interactions \cite{fl,w1,q1}, and
in models used to map biological systems, involving an activator and its
antagonist, the inhibitor \cite{m}.

In this model the first order equations are
\ben
\label{m2f1}
\frac{d\phi}{dx}&=&1-\phi^2+\frac12\,r\,\frac{\chi^2}{\phi^2}+
\frac{s}{\phi^2}-\frac{s}{\phi^4}
\\
\label{m2f2}
\frac{d\chi}{dx}&=&-\,r\,\frac{\chi}{\phi}
\een
We investigate the BPS sector defined by the minima $(\pm1,0)$. Its energy
is given by $(4/3)(1-s)$. In this model, the presence
of interactions leading to negative power in the $\phi$ field makes the
search for defects harder than before, so this example offers another
good illustration of the trial orbit method.

We follow the former steps and try the orbit
$a+b\phi^2+c\phi^4+d\phi^2\chi^2=0$. It gives the good orbit
$\phi^2\chi^2-2s(1-\phi^2)=0$ for $r=1$. We use this orbit in the first
order equation (\ref{m2f1}) to get $d\phi/dx=1-\phi^2$, which is solved
by $\phi(x)=\tanh(x)$. This means that the other field $\chi$ is
\be
\chi(x)=\pm\frac{\sqrt{2s}}{\sinh(x)}
\ee
We see that $\chi(x)$ diverges for $x\to0$, spliting the orbit into
two segments. Because of this singularity we cannot calculate the
corresponding energy, and this indicates that this pair of solutions
is unstable.

The last example is obtained as a two-field generalization of a model
introduced in Ref.~{\cite{cvi99}}, which describes a vacuumless potential.
The potential has a maximum at $\phi=0$, and monotonically
decreases to zero at $\phi\to\infty$. Potentials with such asymptotic
behavior can arise in supersymmetric gauge theories \cite{ads}, and more
recently they have been used in the form of quintessence models
\cite{qui}. They support defects which present quite different
properties and evolution, as compared to the usual defects \cite{cvi99}.
To generalize the model introduced in \cite{cvi99} to the case of two real
scalar fields we follow Ref.~{\cite{b00}}. We introduce the superpotential
\be
W(\phi,\chi)=\arctan[\sinh(\phi)]+{\widetilde W}(\phi,\chi)
\ee
where
\be
{\widetilde W}(\phi,\chi)=r
\frac{\sinh(s\chi)-\sinh(\phi)}{\cosh(\phi)\cosh(s\chi)}
\ee
with $r$ and $s$ as real parameters. The first order equations are
\ben
\frac{d\phi}{dx}&=&\frac1{\cosh(\phi)}-r
\biggl[\frac{1+\sinh(\phi)\sinh(s\chi)}{\cosh^2(\phi)\cosh(s\chi)}\biggr]
\\
\frac{d\chi}{dx}&=&rs\;\biggl[\frac{1+\sinh(\phi)\sinh(s\chi)}{\cosh(\phi)
\cosh^2(s\chi)}\biggr]
\een 
These equations are much harder to solve, but we can use the trial orbit
method to see that the orbit $\phi=s\chi$ is a good
orbit for $r=1/(1+s^2)$. In this case we get the solutions
\be
\phi(x)=s\chi(x)={\rm arcsinh}\biggl[\frac{s^2\;x}{(1+s^2)}\biggr]
\ee
This pair of solutions is of direct interest to high energy physics, as we
can see for instance in \cite{cvi99,cvia}, and in \cite{bbs02} in the case
of quintessence with coupled scalar fields.

The above examples illustrate how efficiently the trial orbit method can be
used to obtain explicit BPS solutions in specific models, and this will
certainly help exploring other systems, involving the yet unknown
possibilities of finding BPS states in models described by three
or more real scalar fields, with discrete symmetry.
 
We thank CAPES, CNPq, PROCAD and PRONEX for financial support. WF thanks
Departamento de Matem\'atica, Universidade Regional do Cariri, 63.100-000
Crato, Cear\'a, Brazil for support, and Funda\c c\~ao Cearense de Amparo
a Pesquisa for a fellowship.

\end{document}